\documentclass[aps,prl,twocolumn,showpacs]{revtex4}
\usepackage{epsfig}

\begin{document}

\title{Long time correlations in Lagrangian dynamics: a key to intermittency in turbulence}
\author{N. Mordant${}^{(1)}$, J. Delour${}^{(2)}$, E. L\'ev\`eque${}^{(1)}$, A. Arn\'eodo${}^{(2)}$ \& J.-F. Pinton${}^{(1)}$}
\affiliation{${}^{(1)}$ Laboratoire  de Physique, \'Ecole Normale SupŽrieure de Lyon, 46 all\'ee d'Italie F-69007 Lyon, France\\
${}^{(2)}$Centre de Recherche Paul Pascal, Avenue Dr. A. Schweitzer, F-33600  Bordeaux, France.}

\begin{abstract}

New aspects of turbulence are uncovered if one considers flow motion from the perspective of a fluid particle (known as the Lagrangian approach) rather than in terms of a velocity field (the Eulerian viewpoint). Using a new experimental technique, based on the scattering of ultrasounds, we have obtained a direct measurement of particle velocities, resolved at all scales, in a fully turbulent flow. It enables us to approach intermittency in turbulence from a dynamical point of view and to analyze the Lagrangian velocity fluctuations in the framework of random walks.  We find experimentally that the elementary steps in the ÔwalkÕ have random uncorrelated directions but a magnitude that is extremely long-range correlated in time. Theoretically, we study a Langevin equation that incorporates these features and we show that the resulting dynamics accounts for the observed one- and two-point statistical properties of the Lagrangian velocity fluctuations. Our approach connects the intermittent  statistical nature of turbulence to the dynamics of the flow.

\end{abstract}


\maketitle
 
Traditional experimental studies of velocity fluctuations in turbulence rely on velocimetry measurement at a fixed point in space. A local velocity probe yields time traces of the velocity fluctuations which are then related to spatial velocity profiles using the Taylor hypothesis~\cite{MonYag}. In this case, the flow is analyzed in terms of the Eulerian velocity field $u(x,t)$. One of the most peculiar feature of homogeneous three-dimensional turbulence is its intermittency, well established in the Eulerian framework~\cite{Frisch}. The statistical properties of the flow depend on the length scale at which it is analyzed. For instance, the functional form of the probability of measuring an Eulerian velocity increment $\Delta_s u(x) = u(x+s)-u(x)$ varies with the magnitude of the length scale $s$. Many studies devoted to the understanding of this feature have been developed along the lines of Kolmogorov and Obhukov 1962 pioneering ideas~\cite{KO62}. In this case, intermittency is analyzed in terms of the anomalous scaling of the moments of the velocity increments in space. It is attributed to the inhomogeneity in space of the turbulent activity and  often analyzed in terms of ad-hoc multiplicative cascade models~\cite{Frisch}. Although very successful at describing the data, these models have failed to connect intermittency with the dynamical equations that govern the motion of the fluid.
Here, we adopt a Lagrangian point of view. It is a natural framework for mixing and transport problems in turbulence~\cite{Ottino}. In addition it has been shown in the passive scalar problem that intermittency is strongly connected to the particular properties of Lagrangian trajectories~\cite{Pumir,Falkovich}.  In the Lagrangian approach, the flow is parameterized by $v(x_0,t)$, the velocity of a fluid particle initially at position $x_0$.  Experimentally, we follow the motion of a single tracer particle and we consider the increments in time of its velocity fluctuations:  $\Delta_\tau v(t) = v(t+\tau)-v(t)$.  Our first observations~\cite{Mordant} have established and described intermittency in this Lagrangian framework. In order to understand its origin, we propose here a new point of view. Since our measurements give access to the individual motion of fluid particles, we study intermittency from a dynamical point of view. We show that the anomalous scaling in the Lagrangian velocity increments traces back to the existence of long-time correlations in the particle accelerations, {\it i.e.} the hydrodynamic forces that drive the particle motion. 

In order to study the motion of Lagrangian tracers, we need to resolve their velocity fluctuations across a wide range of scales. To this end, we use a confined flow with no mean advection, so that fluid particles remain for long times inside a given measurement volume.  The tracking of small tracer particles is achieved using a new acoustic technique based on the principle of a ``continuous Doppler sonar''. The flow volume is continuously insonified with a monochromatic ultrasound which is then scattered by the tracer particle~\cite{Mordant}. This scattered sound is detected by two transducer arrays which yield a measurement of both the particle position, by direct triangulation, and of its velocity, from the Doppler shift.  Indeed, for an incoming sound with frequency $f_0$ , the scattered sound at the receiver has frequency $f(t) = f_0 + \mathbf{k}.\mathbf{v}(t)$ , where $\mathbf{v}(t)$ is the velocity of the tracer particle and $\mathbf{k}$ is the scattering wave vector. This frequency modulation in the acoustic signal is extracted numerically, using a high-resolution parametric method~\cite{JASA}. Figure 1(a) shows the experimental set-up and an example of a particle trajectory; Figure 1(b) gives an example of the time variation of one component of its velocity. A water flow of the von Karman swirling type~\cite{LaPorta,Mordant} is generated inside a cylinder by counter-rotation at $\Omega=7$~Hz of two discs with radius $R=9.5$~cm, fitted with eight blades of height 0.5~cm and set 18~cm apart. The flow power consumption is $\epsilon=25$~W/kg, with velocity fluctuations $u_{rms}=0.98$~m/s. The characteristic size of the velocity gradients is $\ell =(15 u_{rms}^2 / \epsilon )^{1/2} = 880 \; \mu$m, larger than the diameter ($250 \; \mu$m) of the neutrally buoyant tracer particle (density of 1.06). The turbulent Reynolds number of the flow is $R_\ell = u_{rms}\ell/\nu =740$. The large scale flow is axisymmetric and the fluctuations in its center approximate well the conditions of local homogeneous and isotropic turbulence.  The flow is insonified at 2.5 MHz, with the transducers located at the flow wall.  The receiver arrays are placed at 45 degrees on each side of the emission direction. The measurement region is the intersection of the emission and detection cones. The particles act as Lagrangian tracers for times longer than 1~ms (below which inertia cuts off their response), up to times as long as they will stay confined insided the measurement volume, {\it i.e.}  between one and ten $T_L$, the Lagrangian integral time scale (computed from the Lagrangian velocity autocorrelation function). 4000 such events are analyzed, for a total of $1.9\times 10^6$ data points sampled at 6.5 kHz. 

\begin{figure}[h]
    \centering
    \epsfxsize=6cm
    \epsfbox{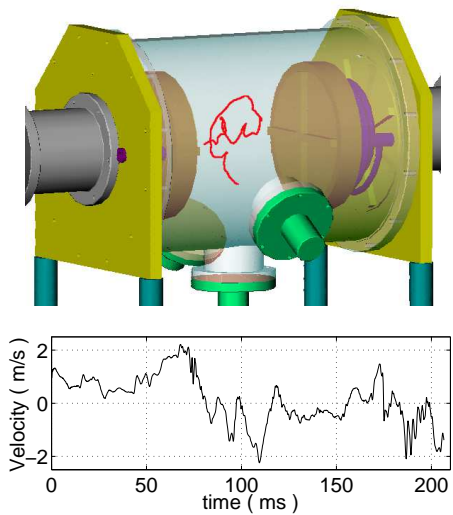}
    \caption{(a) Experimental set-up of the von K\'arm\'an flow, with an example of a particle 3D trajectory, 206 ms ($=9.2 \; T_L$); 
    (b) corresponding velocity variation (one component shown). }
\end{figure}

The probability density functions (PDFs) of the Lagrangian time increments $\Pi_\tau(\Delta v)$, are shown in Fig.2. They are Gaussian at integral scale ($\tau > T_L$) and vary continuously towards the development of stretched exponential tails as the time increments decrease towards the dissipative time scale~\cite{Yeung,Mordant}. The outer curve in Fig.2 is the PDFs of Lagrangian acceleration measured by LaPorta {\it et al.}~\cite{LaPorta} in the same flow geometry and at a comparable turbulent Reynolds number. 
\begin{figure}[h]
    \centering
    \epsfxsize=6cm
    \epsfbox{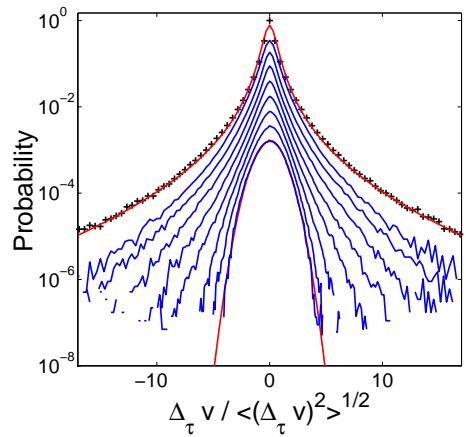}
    \caption{PDF $\Pi_{\tau}(\Delta v)$ calculated for time lags $\tau =1.3, 2.7, 5.4, 11.2, 22.4, 44, 89.3$ and 174\% of $T_L$.  The curves are computed for variations normalized to the standard deviation and displayed with vertical shift for clarity.  The underlying crosses in the outer curve  correspond to the MRW model --- see text ---  with $\lambda^2=0.115$.}
\end{figure}
This evolution of the PDFs $\Pi_\tau(\Delta v)$ leads to an anomalous scaling of the velocity structure functions $\langle | \Delta_\tau v |^q \rangle \sim \tau^{\zeta(q)}$, with $\zeta(q)$ a non-linear function of $q$. The intermittency spectrum is usually measured experimentally using the ESS ansatz~\cite{Benzi}, {\it i.e.} via the relative scaling $\langle | \Delta_\tau v |^q \rangle \sim \langle | \Delta_\tau v |^2 \rangle^{\xi(q)}$, where $\xi(q) = \zeta(q)/\zeta(2)$. In Kolmogorov K41 phenomenology~\cite{Frisch}, the Lagrangian second order structure function is assumed to scale linearly with the time increment $\tau$ so that $\zeta(2)$ is considered as being equal to one, eventhough it is quite hard to establish it experimentally~\cite{Mordant}. Given the statistics available in our experiments, one is limited to moments up to order 6. We observe that $\xi(q)$ is well represented by a quadratic law~\cite{Mordant}:
\begin{equation}
\xi(q) =  (1/2 + \lambda_L^2)q -  \lambda_L^2 q^2 / 2 , \;  \; \lambda_L = 0.115 \pm 0.01 \ .
\end{equation}
Note that the Lagrangian value of the intermittency parameter  ($\lambda_{L}^2= 0.115$) is larger than the value ($\lambda_{E}^2 = 0.025$), measured  from hot-wire anemometry or in direct numerical simulations~\cite{Delour}.  This is expected~\cite{Borgas}, and mainly comes from the facts that time increments are measured here and that the reference structure function is the second order one.

Intermittency is thus observed and quantified in both Lagrangian and Eulerian frameworks.  In contrast to traditionnal Eulerian studies where intermittency is described in terms of multiplicative processes, we look here for a dynamical origin.  We consider the statistics of the fluid particleÕs fluctuating velocity in analogy with a random walk.  We write a velocity increment over a time lag  $\tau$ as the sum of contributions over small times  $\tau_1$: 
\begin{equation}
\Delta_\tau v(t) = v(t+\tau) - v(t) = \sum_{n=1}^{\tau/\tau_1} \Delta_{\tau_1} v(t+n\tau_1) \ .
\end{equation}
If the incremental ``steps''   of duration $\tau_1$ were independent (and identically distributed), the PDF $\Pi_\tau(\Delta v)$  would readily be obtained as a convolution of the elementary distribution at scale $\tau_1$, $\Pi_{\tau_1} (\Delta v)$ --- plus an eventual convolution kernel to account for stationarity at large scales. Such a regular convolution process corresponds to the Kolmogorov K41 picture of turbulence~\cite{Frisch}; the particle velocity fluctuations are Brownian and the scaling is monofractal.  A first important result of our analysis is to show that the elementary steps are not independent. The auto-correlation of the signed increments, $\Delta_{\tau_1} v(t)$, decays very rapidly (cf. Fig.3):  the correlation coefficient drops under 0.05 for time separations larger than $2 \tau_1$. However, if one considers the amplitude of the ``steps'' ($| \Delta_{\tau_1}v(t) |$),  one finds that the auto-correlation decays very slowly and only vanishes at the largest time scales of the turbulent motion. Recast in terms of the random walk, our results show that the amplitudes of the ``steps''  are long-range correlated in time although their directions are not. As this point is fundamental for our approach, we have verified it using a Lagrangian tracking algorithm in a Direct Numerical Simulation (DNS) of the Navier-Stokes equations, using a pseudo-spectral solver, at $R_\ell = 75$, and for the same ratio $\tau_1/T_L$ --- see inset of Fig.3. The results are in remarkable agreement with our measurements. All increments are correlated for $\Delta t<\tau_1$, the time over which they are computed. Above $\tau_1$, the correlation of the signed increments rapidly drops while the correlation coefficient of their absolute values decays very slowly, to vanish only for $\Delta t  > 3 \; T_L$. This behavior is observed for $\tau_1$ chosen from the smallest resolved time scales to inertial range values.  These observations also persist in the limit of very small time increments, and thus presumably for the acceleration of the fluid particle and thus for the forces acting on it. 

\begin{figure}[h]
    \centering
    \epsfxsize=6cm
    \epsfbox{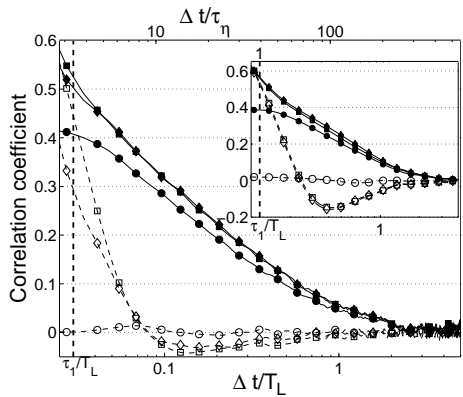}
    \caption{Variation of the normalized correlation coefficient $ \chi(f,g)(\Delta t)=\langle ( f(t + \Delta t)- \langle f \rangle )( g(t)- \langle g \rangle ) \rangle  / \sigma_f \sigma_g $. Two velocity components are considered: the squares and diamonds mark the $\chi ( \Delta_{\tau_1}v_x, \Delta_{\tau_1}v_x)$ and $\chi ( \Delta_{\tau_1}v_y, \Delta_{\tau_1}v_y)$ auto-correlation functions while the circles mark the cross-correlation $\chi ( \Delta_{\tau_1}v_x, \Delta_{\tau_1}v_y)$; curves with filled symbols are computed using the absolute value of the increments while the curves  with open symbols are computed using the full signed increments. The main curve corresponds to the experiment at $R_\ell = 740$, with $\tau_1 = 0.03 \;  T_L$. The inset shows similar results for a Direct Numerical Simulation at $R_\ell = 75$.  }
\end{figure}

Theoretically, one would like to understand this behavior from the hydrodynamic forces in the Navier-Stokes equations. Such a direct analytical treatment is out of reach at present. We propose, as a natural first step, to study a surrogate dynamical equation of the Langevin type. In this  procedure, one considers a one-dimensional variable,  $W(t)$, representing the particle velocity, driven by a stochastic force. If this force is chosen as a white noise then $W(t)$ has the dynamics of Brownian motion: its statistics is monofractal with a similarity exponent equal to 1/2 --- the increments scale as $\langle | W(t+\tau)-W(t) |^p \rangle_t \sim \tau^{p/2}$, corresponding to the non-intermittent Kolmogorov 1941 picture. In order to account for intermittency, one needs to ascribe other properties to the stochastic force. Guided by our experimental results, we build a stochastic force having a random direction and a long-range correlation in its magnitude. Specifically,  its direction is modeled by a Gaussian variable $G(t)$, chosen white in time, with zero mean and unit variance. The amplitude of the force, $A(t)$, being a positive variable, is written $A(t) = \exp[\omega(t)]$  where the magnitude $\omega(t)$ is a stochastic process that satisfies:
\begin{equation}
\langle  \omega(t) \omega(t+\Delta t) \rangle_t  =  - \lambda^2  \ln(\Delta t/ T_L)   \;  {\rm for} \;  \Delta t  < T_ L,
\label{eqmrw} 
\end{equation}
and $0$ otherwise --- $\lambda^2$ being an adjustable parameter.  When discretized, this dynamics corresponds to a one-dimensional Multifractal Random Walk (MRW)~\cite{Bacry}. Analytical calculations show that the resulting dynamical variable $W(t)$ has multi-scaling properties. The moments  have scaling laws, $\langle | \Delta_\tau W |^q \rangle \sim  \tau^{\zeta(q)}$, with $\zeta(q) = (1/2 + \lambda^2)q -  \lambda^2 q^2 / 2$, so that  $\lambda^2$ in equation~(\ref{eqmrw}) is the intermittency parameter of the model~\cite{Bacry}. It is a fundamental point that the same parameter $\lambda^2$ governs both the evolution of the PDFs of the increments (one-time statistics) and the time correlation of the process (two-time statistics). 

\begin{figure}[h]
    \centering
    \epsfxsize=5.5cm   \epsfbox{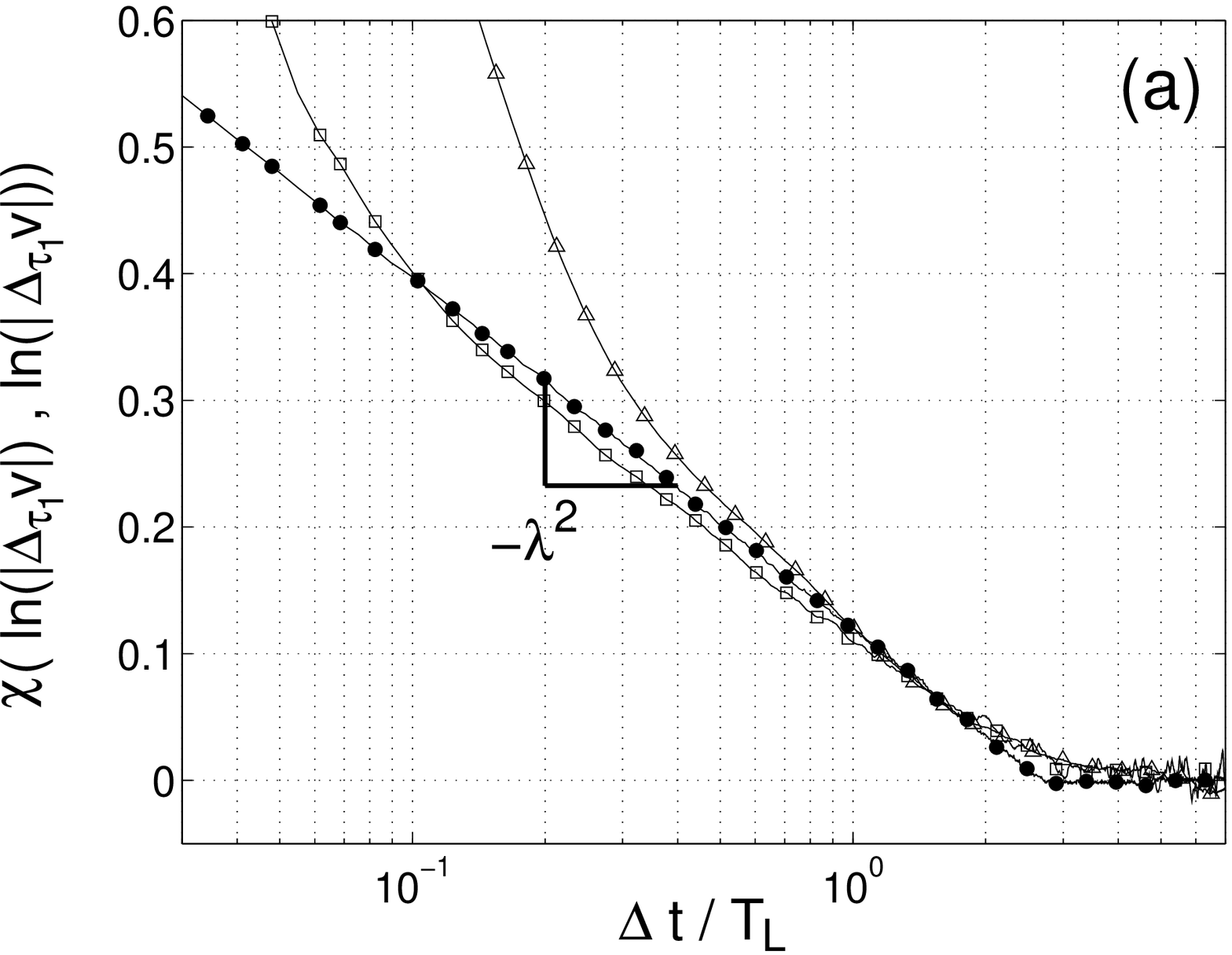} \\
    \epsfxsize=5.5cm   \epsfbox{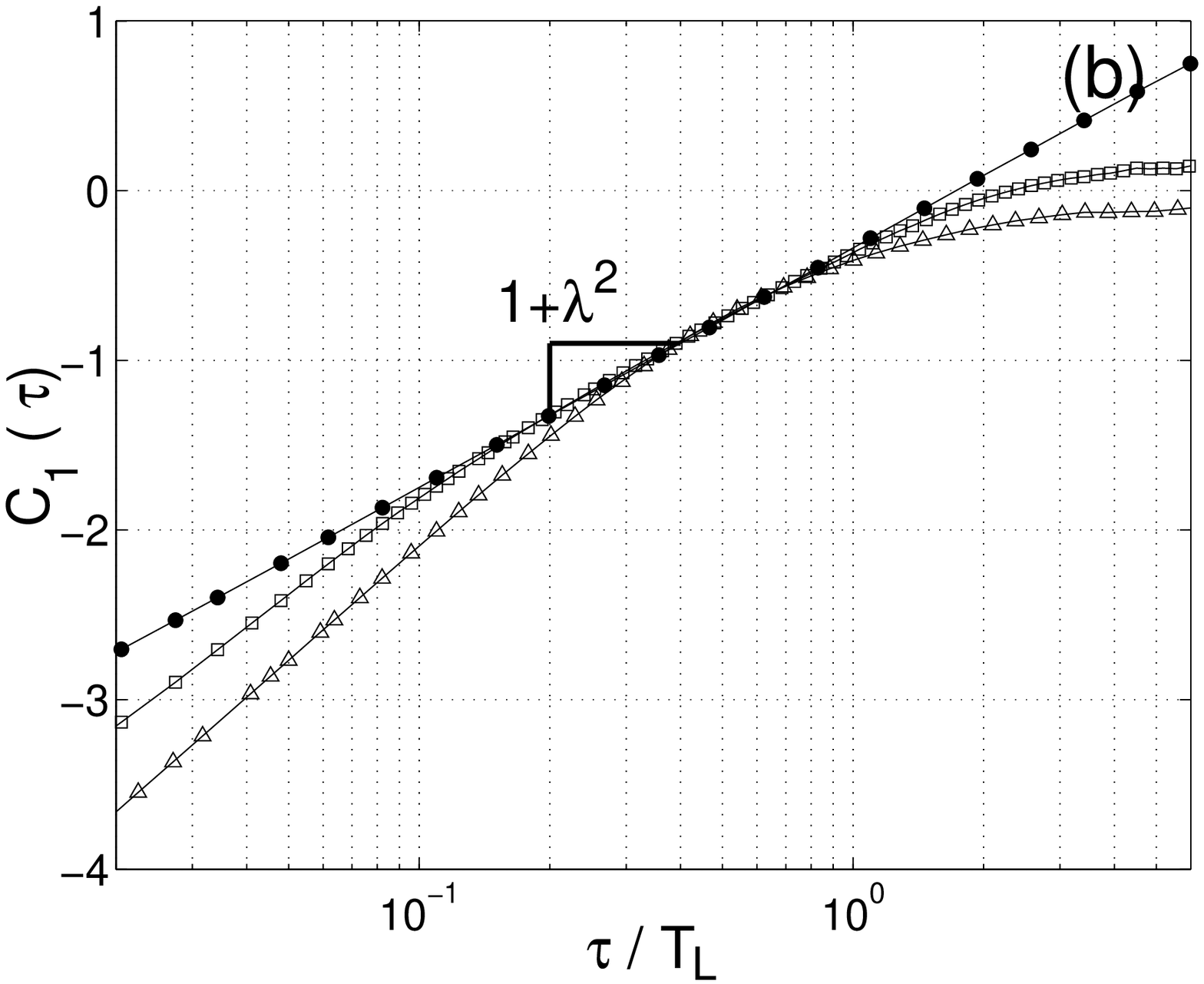} \\
    \epsfxsize=5.5cm   \epsfbox{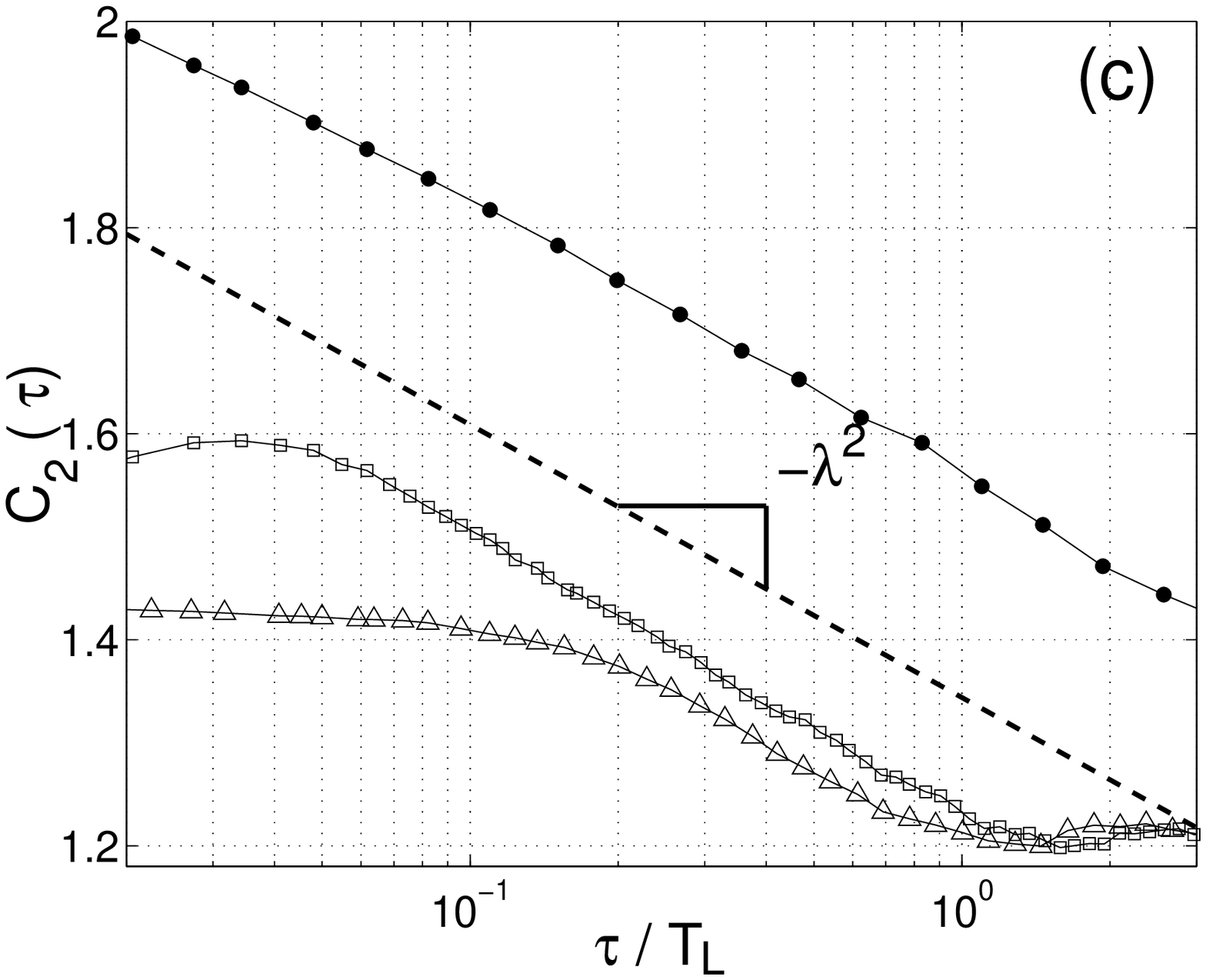} \\
    \caption{Experimental (open squares) and numerical data (open triangles) compared to the predictions of the MRW model (filled circles).  (a) Correlation in time of the magnitude of one component of the velocity increments, $\chi(\log|\Delta_{\tau_1} v_x |, \log|\Delta_{\tau_1} v_x |)$, computed for a time lag $\tau_1 = 0.03 T_L$.  (b) \& (c): first and second order cumulants versus time scale $\tau$.}
\end{figure}

We show that this model captures the essential features of the Lagrangian data. First, in order to test the relevance of equation~\ref{eqmrw}, we have computed, from the experimental and numerical data, the auto-correlation function of the logarithm of the amplitude of infinitesimal Lagrangian  velocity increments: $\chi(\log|\Delta_{\tau_1} v_x |, \log|\Delta_{\tau_1} v_x |)$. Figure 4(a) confirms that the logarithmic decrease build in the MRW model (equation~\ref{eqmrw}) is observed both in the experimental and numerical data; it yields the estimate $\lambda^2= 0.115 \pm 0.01$. Second, we check the relevance of the model for the description of the one-time statistics of the Lagrangian increments $\Delta_\tau v$. We note --- Fig.2, upper curve--- that the choice $\lambda^2= 0.115$ yields a PDF for the stochastic force that is in remarquable agreement with experimental measurements of fluid particle accelerations~\cite{LaPorta}. The agreement at larger time scales is evidenced on the behavior of the first two cumulants. Cumulants are computed with more reliability that the moments and are related to them through $\langle | \Delta_\tau v |^q \rangle = \langle \exp \left( q \ln | \Delta_\tau v | \right) \rangle = \exp \left( \sum_n C_n(\tau) q^n / q! \right)$.  In the MRW model, one can analytically derive~\cite{Bacry}: 
\begin{equation}
C_1(\tau)=(1+\lambda^2) \ln(\tau) , \;      C_2(\tau)=-\lambda^2 \ln(\tau),   
\label{eqcum}
\end{equation}
all higher order cumulants being null. $C_1(\tau)$ and $C_2(\tau)$ computed from the experimental and numerical data are shown in Figures 4(b) and 4(c) and compared to MRW model predictions when the intermittency parameter is set to the value $\lambda^2= 0.115$ that is derived from the correlations in the dynamics. One observes that in each case the agreement is excellent; the slope of the variation $\partial C_{1,2}(\tau)/\partial \ln \tau$ in the inertial range is correctly given by equation~(\ref{eqcum}). The same intermittency parameter thus governs the anomalous scaling of the Lagrangian velocity increments and their long-time dynamical correlations.

We therefore believe that long-time correlations in the Lagrangian dynamics are a key feature for the understanding of intermittency, which leads to a new dynamical picture of turbulence. Long-time correlations and the occurrence of very large fluctuations at small-scales dominate the motion of a fluid particle. It can be understood if, along its trajectory, the particle encounters very intense small-scale structures (vortices and stagnation points) over a more quiet background. Intermittency is then due to the nature and distribution of these small scale structures. Indeed, the analogy with a random walk suggests that the statistics at all scales can be recovered if one ascribes two properties to the small scales: 1) the probability density function of fluid particle accelerations and 2) the functional form of their time correlations. In the Lagrangian framework, these features are directly linked to the Navier-Stokes equations that govern the elementary changes in the velocity (momentum) of the fluid particles. It thus gives a possibility to derive intermittency from the constitutive physical equations. Although this may be quite a theoretical challenge, direct numerical simulations look promising as they allow the study of the flow dynamical fields (pressure, velocity gradient tensor, etc.) along the trajectory of individual fluid particles.

{\bf Acknowledgements.} This work is supported by the French Ministre de La Recherche (ACI), and the Centre National de la Recherche Scientifique under GDR Turbulence.  Numerical simulations are performed at CINES (France) using an IBM SP computer. We thank P. Chanais, O. Michel, B. Portelli, P. Holdsworth for fruitful discussions and we gratefully acknowledge the help of P. Metz, M. Moulin and L. de Lastelle.

\end{document}